\documentclass[manuscript,screen]{acmart}
\AtBeginDocument{%
        \providecommand\BibTeX{{%
                        \normalfont B\kern-0.5em{\scshape i\kern-0.25em b}\kern-0.8em\TeX}}}
\usepackage[noend]{algpseudocode}
\usepackage{graphicx}
\usepackage{caption}
\usepackage{subcaption}
\usepackage{amsmath}
\usepackage{todonotes}
\usepackage{xspace}
\usepackage{microtype}
\usepackage{color}
\usepackage{ifthen}
\usepackage{pgf,tikz,pgfplots}
\usepackage{tikz-3dplot}
\usepackage{tkz-graph}
\usepackage{scalerel}
\usepackage{algorithmicx}
\usepackage{algorithm}
\usepackage{algpseudocode}
\usepackage{comment}
\usepackage{float}
\usepackage{placeins}
\usepackage{longtable}

\usepackage[tableposition=below]{caption}
\newtheorem{theorem}{Theorem}

\newcommand{\AR}{\mbox{\emph{AR}}}
\newcommand{\hull}{\mbox{\emph{conv}}}

\newcommand{\ignore}[1]{}
\newcommand{\old}[1]{{}}

\date{}
\usetikzlibrary{calc,quotes,angles}

\begin{document}

        \title{Area-Optimal Simple Polygonalizations: The CG Challenge 2019}
        \author{Erik D.~Demaine}
        \email{edemaine@mit.edu}
        \orcid{0000-0003-3803-5703}
        \affiliation{
                \institution{CSAIL, MIT}
                \city{Cambridge, MA}
                \country{USA}
        }
        \author{S\'andor P.~Fekete}%
        \email{s.fekete@tu-bs.de}%
        \orcid{0000-0002-9062-4241}
        \affiliation{
                \institution{{Department} of Computer Science, TU Braunschweig}
                \city{Braunschweig}
                \country{Germany}
        }
        \author{Phillip Keldenich}
        \email{p.keldenich@tu-bs.de}
        \orcid{0000-0002-6677-5090}
        \affiliation{
                \institution{{Department} of Computer Science, TU Braunschweig}
                \city{Braunschweig}
                \country{Germany}
        }
        \author{Dominik Krupke}
        \email{d.krupke@tu-bs.de}
        \orcid{0000-0003-1573-3496}
        \affiliation{
                \institution{{Department} of Computer Science, TU Braunschweig}
                \city{Braunschweig}
                \country{Germany}
        }
        \author{Joseph S.~B.~Mitchell}
        \email{joseph.mitchell@stonybrook.edu}
        \orcid{0000-0002-0152-2279}
        \affiliation{
                \institution{{Department} of Applied Mathematics and Statistics, Stony Brook University}
                \city{Stony Brook, NY}
                \country{USA}
        }

%

\setcopyright{acmcopyright}

\acmJournal{JEA}
\acmVolume{99}
\acmNumber{99}
\acmArticle{99}
\acmMonth{13}

\ccsdesc[500]{Theory of computation~Design and analysis of algorithms}
\ccsdesc[300]{Theory of computation~Computational Geometry}

        \begin{abstract}
        We give an overview of theoretical and practical aspects
	of finding a simple polygon of minimum (\textsc{Min-Area}) or maximum
        (\textsc{Max-Area}) possible area for a given set of $n$ points in the plane. 
	Both problems are known to be $\mathcal{NP}$-hard and 
	were the subject of the 2019 Computational Geometry Challenge, 
	which presented the quest of finding good solutions to more than
	200 instances, ranging from $n=10$ all the way to $n=1,000,000$.
        \end{abstract}

\maketitle

\section{Introduction}

\subsection{The Computational Geometry Challenge}
The ``CG:SHOP Challenge'' (Computational Geometry: Solving Hard
Optimization Problems) originated as a workshop at the 2019
Computational Geometry Week (CG Week) in Portland, Oregon in June,
2019.  The goal was to conduct a computational challenge competition
that focused attention on a specific hard geometric optimization
problem, encouraging researchers to devise and implement solution
methods that could be compared scientifically based on how well they
performed on a database of instances. While much of computational
geometry research has targeted theoretical research, often seeking
provable approximation algorithms for $\mathcal{NP}$-hard optimization problems,
the goal of the CG Challenge was to set the metric of success based on
computational results on a specific set of benchmark geometric
instances. The 2019 CG Challenge focused on the problem of computing
simple polygons whose vertices were a given set of points in the
plane. 

A tangible outcome is a new type of special issue, presented in this volume: 
a series of papers focusing on algorithm engineering methods for \emph{one} 
difficult optimization problem. 
In this survey, we provide background and foundations
of the underlying problem and give an overview of the results
and contributions. 

\subsection{The 2019 Challenge Problem: Polygonizations of Optimal Area}
The Traveling Salesman Problem (TSP) is one of the classic optimization problems:
For a given set of locations and pairwise distances,
find a shortest roundtrip that visits each position exactly once
and returns to the start. 
In a geometric setting, in which locations correspond to a given 
set $P$ of $n$ points in the plane, and distances between points
are induced by the Euclidean metric, it is a straighforward consequence
of the triangle inequality that an optimal tour corresponds to a simple
polygon $\mathcal{P}$ with vertex set $P$, such that $\mathcal{P}$ has minimum
total perimeter length.

This geometric motivation makes it natural to consider simple polygons with a
given set of vertices that minimize another basic geometric measure: the
enclosed area. 
This was considered in the past in the context of surface
reconstruction, e.g., by O'Rourke~\cite{o1980polyhedral}; as sketched in
Section~\ref{sec:pick}, there is also a close connection to point separation,
which has gained importance in Artificial Intelligence. The same context has
also raised interest in the corresponding maximization problem: For a given
set $P$ of $n$ points in the plane, find a simple polygon $\mathcal{P}$
with vertex set $P$ of maximum possible area.
In the following, we refer to these two problems as \textsc{Min-Area} and
\textsc{Max-Area}, respectively.

\medskip
\textbf{Problem:} \textsc{Min-Area} and \textsc{Max-Area}

\textbf{Given:} A set $P$ of $n$ points in the plane.

\textbf{Goal:} A simple polygon with vertex set $P$ of minimum (\textsc{Min-Area})
or maximum (\textsc{Max-Area}) possible enclosed area. 

\begin{figure}[h]
        \centering
        \includegraphics[height=0.3\textwidth]{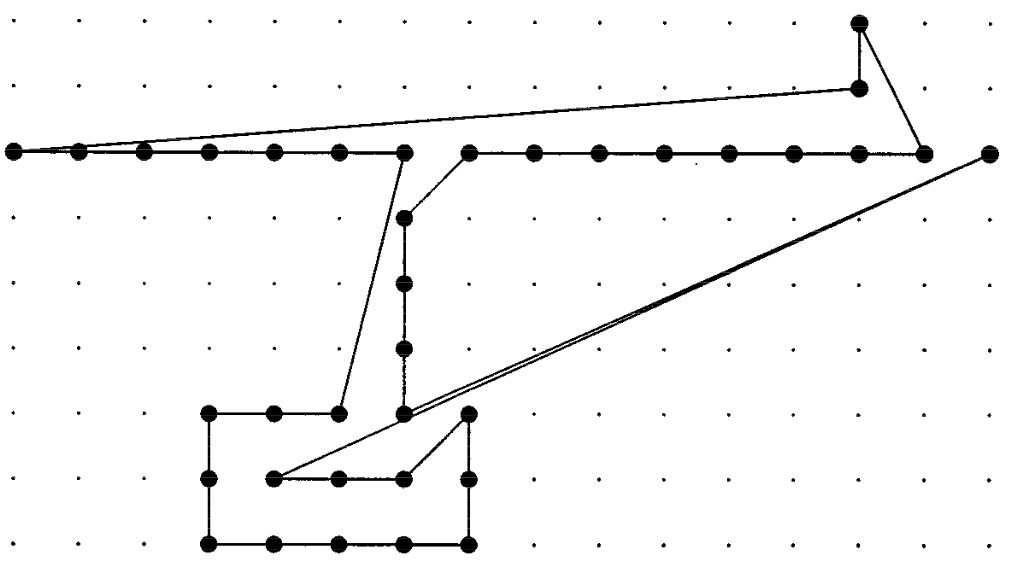}
        \caption{A point set $P$ and its only minimum-area polygon $\mathcal{P}$, containing several long edges. (Example from~\cite{f-gtsp-92,fekete2000simple}.)}
        \label{fig:badexample}
\end{figure}

\medskip
There are a number of features that make these problems suitable for 
optimization challenges, based on notable similarities to and differences
from the TSP. As shown by Fekete~\cite{f-gtsp-92,fekete1993area,fekete2000simple}, 
both \textsc{Min-Area} and
\textsc{Max-Area} are $\mathcal{NP}$-hard; however, while membership in $\mathcal{NP}$
for the Euclidean
TSP is a long-standing, 
famous open problem (e.g., Problem \#33 in The Open Problems Project~\cite{topp}), it is
straightforward for area optimization, so issues of numerical
stability and checking vailidity of solutions do not come into play. 
On the other hand, edges in a polygon
with small area need not be short, as shown in Figure~\ref{fig:badexample}. 
This makes it challenging to restrict potential neighbors of a point in a 
good polygonization, increasing the difficulty of employing local search methods for efficient 
algorithms. This explains an apparent practical distinction to the 
TSP: While provably optimal solutions for TSP benchmark instances 
of considerable size have been known for a while, exact methods
for area optimization appear more elusive.

\subsection{Related Work}

\subsubsection{History and Background}
The origins of the problem of finding 
area-optimal polygons can be traced back at least to the early days
of Computational Geometry (O'Rourke~\cite{o1980polyhedral} in 1980).
Resolving the complexity was posed by Suri in 1989 at CCCG, and restated in a different
context by Mitchell \cite{mitchell1993approximation} and Mitchell and Suri~\cite{jsb1995separation}. 

Studying the set of possible polygonizations is of great interest in various
applications~\cite{aichholzer2002enumerating,zhu1996generating,
ackerman2009improved,abellanas1993updating,arkin2003reflexivity}. 
Auer and Held~\cite{auer1996rpg} gave methods for 
generating random polygonizations.
O'Rourke and Suri~\cite{rourke30polygons} raised the question of estimating
the number of polygonizations as a function of the number $n$ of points. 
At this point, this is known to lie between $4.642^n$ \cite{garcia2000lower} 
and $56^n$ \cite{sharir2013counting}. 

\subsubsection{Complexity}
Fekete \cite{fekete2000simple, f-gtsp-92} gave a proof of
$\mathcal{NP}$-completeness for both problems, based on a reduction
from \textsc{Hamiltonicity of Planar Cubic Directed Graphs},
and generalized this proof to higher dimensions:
For any fixed $1 \leq k \leq d$ and
$2 \leq d$, it is $\mathcal{NP}$-hard to find the polyhedron with
$k$-dimensional faces of minimum total volume
for given vertices in $d$-dimensional space.
He
also showed that no {polynomial-time
approximation scheme} (PTAS) exists for \textsc{Min-Area}
and presented
a $\frac{1}{2}$-approximation algorithm for \textsc{Max-Area}
\cite{fekete1993area,f-gtsp-92}. Moreover, he proved that
the $\mathcal{NP}$-hardness of the minimization problem also applies for higher
dimensions. More specifically he showed that for given $1 \leq k \leq d$ and
$2 \leq d$ it is $\mathcal{NP}$-hard to find the minimal volume polyhedron with
$k$-dimensional faces for given vertices. 

\subsubsection{Heuristics}
Recent work was mainly focused on
finding new heuristics for both \textsc{Min-Area} and \textsc{Max-Area}. 
Taranilla et al.~\cite{taranilla2011approaching} proposed three different heuristics.
Peethambaran et al.~\cite{peethambaran2016empirical,peethambaran2015randomized} 
proposed randomized and greedy algorithms for \textsc{Min-Area} and 
$d$-dimensional variants of both problems.

\subsubsection{Other Challenges}
The Open Problems Project (TOPP), maintained by Demaine, Mitchell and O'Rourke~\cite{topp},
is a library of long-standing unsolved problems. On the more
practical side, there have been different efforts, based on benchmark libraries,
such as the TSPLIB~\cite{reinelt1991tsplib}. Since 1990, the DIMACS
implementation challenges have addressed questions of determining realistic
algorithm performance where worst-case analysis is overly pessimistic and
probabilistic models are too unrealistic. Since 1994, the Graph Drawing (GD)
community has held annual contests in conjunction with its annual symposium to
monitor and challenge the current state of the graph-drawing technology and to
stimulate new research directions for graph layout algorithm. 
More recently, a variety of implementation challenges
have gained traction in the world of programming and optimization,
but not yet in the field of
Computational Geometry.

%
\old{ Joe's version:
The ``CG:SHOP Challenge'' (Computational Geometry: Solving Hard
Optimization Problems) originated as a workshop at the 2019
Computational Geometry Week (CG Week) in Portland, Oregon in June,
2019.  The goal was to conduct a computational challenge competition
that focussed attention on a specific hard geometric optimization
problem, encouraging researchers to devise and implement solution
methods that could be compared scientifically based on how well they
performed on a database of instances. While much of computational
geometry research has targeted theoretical research, often seeking
provable approximation algorithms for $\mathcal{NP}$-hard optimization problems,
the goal of the CG Challenge was to set the metric of success based on
computational results on a specific set of benchmark geometric
instances. The 2019 CG Challenge focussed on the problem of computing
minimum-area polygons whose vertices were a given set of points in the
plane.  This Challenge generated a strong response from many research
groups, from both the computational geometry and the combinatorial
optimization communities, and resulted in a lively exchange of
solution ideas.
}

\subsection{Outcomes}
The contest generated a large number of contributions, both for
\textsc{Min-Area} and \textsc{Max-Area}. In the aftermath, the
top teams were invited to describe their methods in detailed
papers, which form the substance of this special issue.

\begin{itemize}
\item Julien Lepagnot, Laurent Moalic, Dominique Schmitt: 
Optimal area polygonization by triangulation and ray-tracing~\cite{area-omega}

\item Lo\"ic Crombez, Guilherme D. da Fonseca, Yan Gerard:
Greedy and Local Search Solutions to the Minimum and Maximum Are~\cite{area-crombez}

\item Nir Goren, Efi Fogel, Dan Halperin: 
Area-optimal polygonization using simulated annealing~\cite{area-tau}

\item G\"unther Eder, Martin Held, Steinpor Jasonarson, Philipp Mayer, Peter Palfrader:
2-Opt moves and flips for area-optimal polygonizations~\cite{area-salzburg}

\item Natanael Ramos, Rai Caetan de Jesus, Pedro de Rezende, Cid de Souza, Fabio Luiz Usberti: 
Heuristics for area optimal polygonizations~\cite{area-campinas}
\end{itemize}

In addition, there is one paper focusing on exact methods for computing provably 
optimal solutions.

\begin{itemize}
\item S\'andor P. Fekete, Andreas Haas, Phillip Keldenich,  Michael Perk, Arne Schmidt:
Computing area-optimal simple polygonization~\cite{area-exact}
\end{itemize}

In the rest of this survey paper, we provide a discussion of specific aspects
of mathematical connections between area optimization and grid points (Section~\ref{sec:pick}),
approximation algorithms (Section~\ref{sec:approx}), and an overview of contest 
results (Section~\ref{sec:contest}).

\section{Pick's Theorem and Integrality}
\label{sec:pick}

For a simple polygon $\mathcal{P}$ with $n$ vertices, computing the 
Euclidean length of its perimeter involves evaluating a sum of square roots,
for which membership in $\mathcal{NP}$ is a long-standing open problem~\cite{topp}.
This differs from computing the area of $\mathcal{P}$,
which can be evaluated quite efficiently, e.\,g., see O'Rourke~\cite{o1998computational}.
An elegant combinatorial answer is given by
Pick's theorem. (This also implies benign objective values,
in particular for vertices whose coordinates are all even numbers,
as chosen in the contest.)

\begin{theorem}[Pick~\cite{pick1899geometrisches}]
\label{Pick's}
Let $\mathcal{P}$ be a simple polygon with integer vertices; let
$i(\mathcal{P})$
be the number of grid points contained in the interior of
$\mathcal{P}$, and let $b(\mathcal{P})$ be the number of grid points on the
boundary of $\mathcal{P}$. Then
\[\mbox{AR}(\mathcal{P})=\frac{1}{2}b(\mathcal{P})+i(\mathcal{P})-1.\]
\end{theorem}

\begin{figure}[h]
        \centering
        \includegraphics[height=0.5\textwidth]{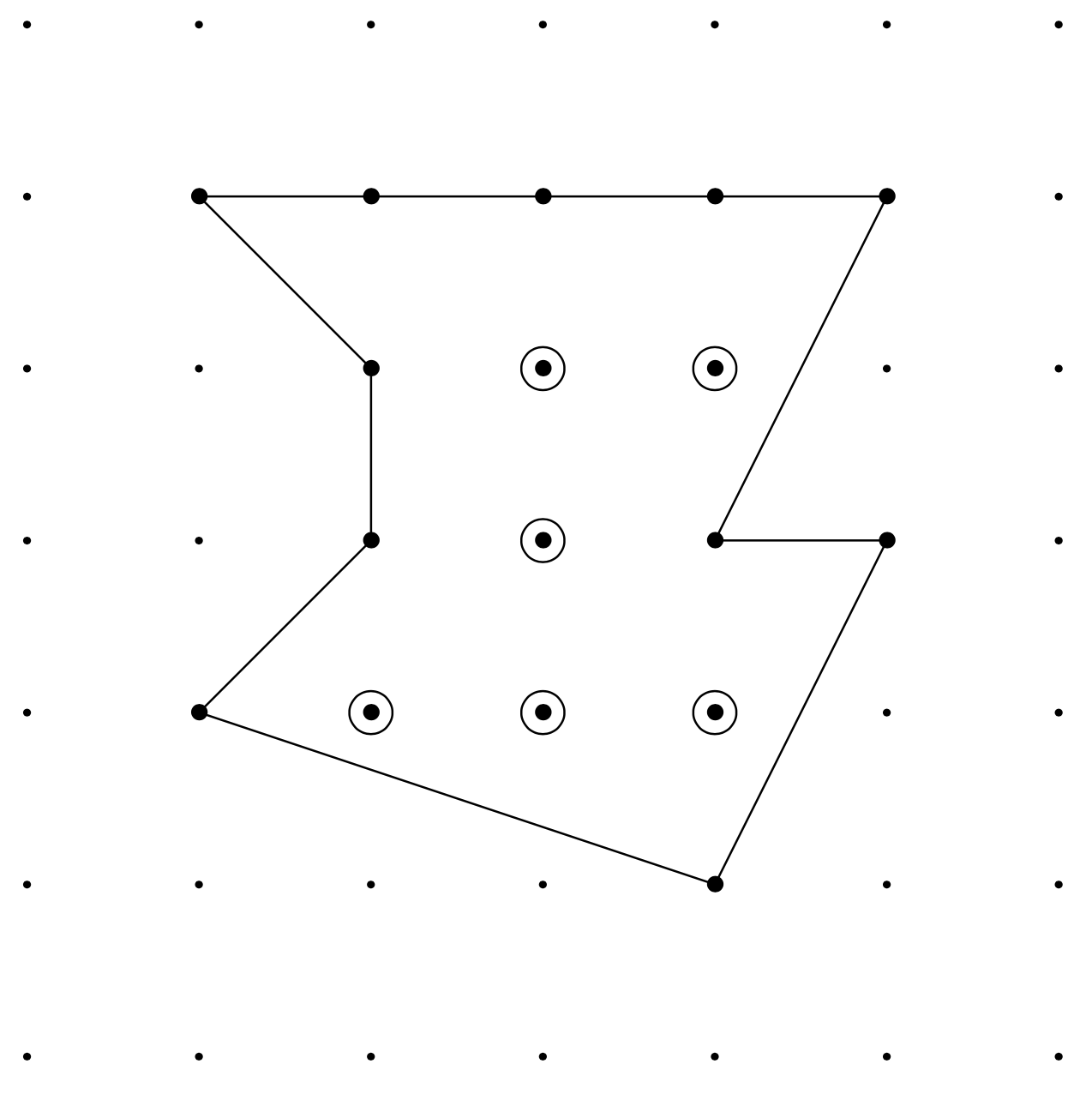}
        \caption{Pick's theorem: A simple grid polygon $\mathcal{P}$
with $b(\mathcal{P})$ grid points on its boundary and $i(\mathcal{P})$
grid points in its interior has area $\frac{1}{2}b(\mathcal{P})+i(\mathcal{P})-1$. (Here shown
for $b(\mathcal{P})=11$ and $i(\mathcal{P})=6$.)}
        \label{fig:pick}
\end{figure}

There are several elegant ways to prove
Pick's theorem, three of which can be found in \cite{funkenbusch1974euler,
coxeter1969introduction, gaskell1976triangulations}. 
For a discussion of alternative approaches see the article by Niven and Zuckermann~\cite{niven1967lattice}.
There are numerous generalizations to other than the orthogonal grid, e.\,g., by Ren and
Reay~\cite{ren1987boundary}; see Reeve~\cite{reeve1957volume} for a generalization
to higher dimensions.

Pick's theorem also provides a bridge to issues of point separation:
Maximizing the enclosed area amounts to finding a simple polygon
that captures as many additional grid points as possible,
while minimizing the area corresponds to excluding as many as possible.
As the $n$ given points must lie on the boundary of $\mathcal{P}$, and only
points within the convex hull of $P$ come into play, we get the following
lower and upper bounds for the area.

\begin{theorem}
\label{homeward}
Let $P$ be a set of $n$ points in the plane that all have integer coordinates.
Let $h_i(P)$ denote the number of points of the integer grid that are not
contained in $P$ and strictly inside the convex hull, and let $h_b(P)$ be the
number of grid points not in $P$ that are on the boundary of the convex hull.

Then for any simple polygon $\mathcal{P}$ on the vertex set $P$, we have
\[\frac{n}{2}-1\leq \AR(\mathcal{P})\leq  \frac{n}{2}+\frac{h_b(P)}{2}+h_i(P)-1.\]
\end{theorem}

\begin{figure}[h]
        \centering
        \includegraphics[height=0.5\textwidth]{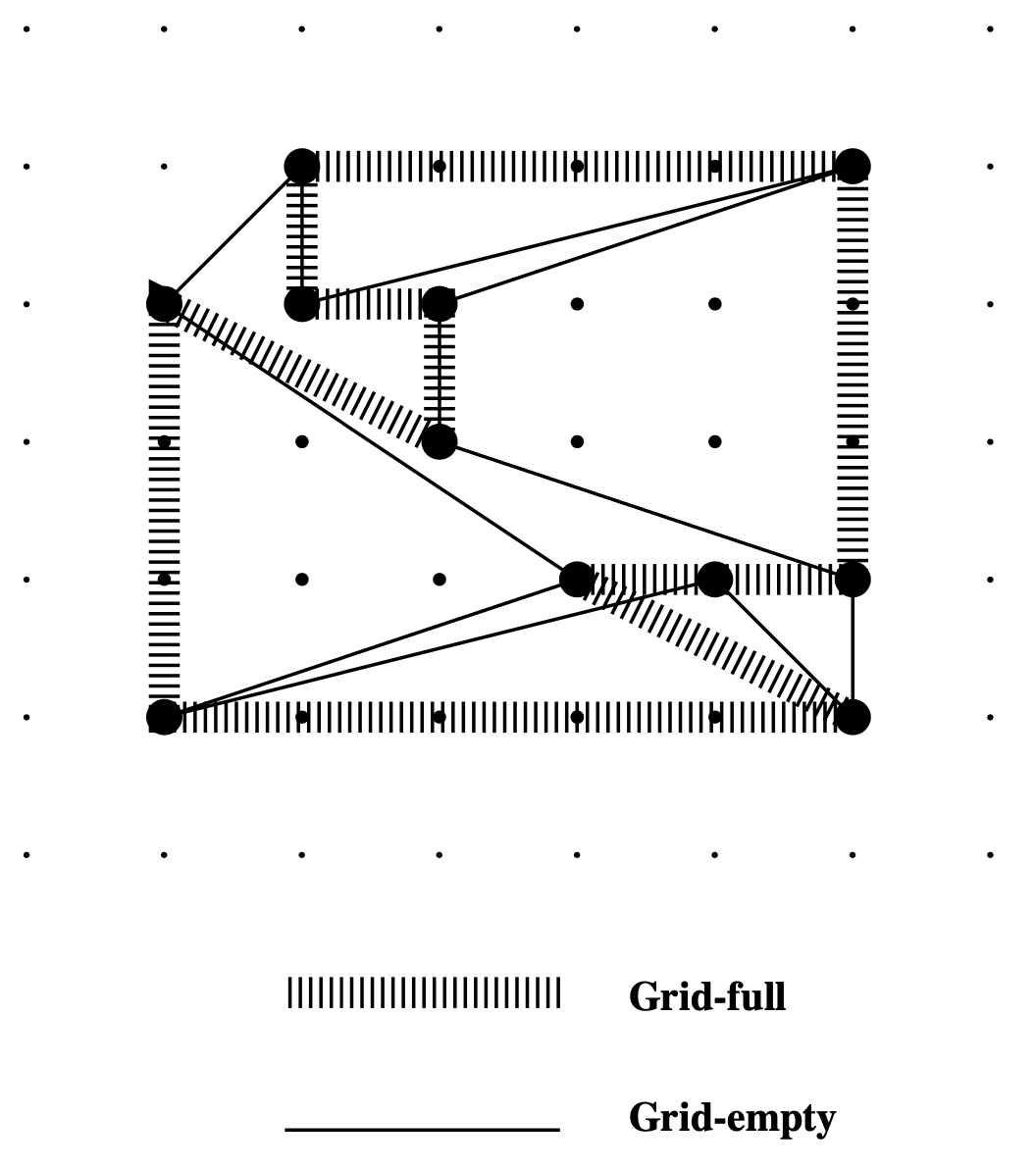}
        \caption{Lower and upper bounds on polygon area, implied by Pick's theorem: 
For a set $P$ of grid points (bold), shown a \emph{grid-empty} simple grid polygon that contains only the given points
and a \emph{grid-full} polygon that contains all grid points of the convex hull to the
maximum possible extent.}
        \label{fig:gridbounds}
\end{figure}

See Figure~\ref{fig:gridbounds} for an illustration.
Deciding whether either of these bounds can be met is 
already $\mathcal{NP}$-complete~\cite{fekete2000simple}.
At the same time, they motivate using the area of the convex hull
as a reference, which was used in the contest. 

\section{Approximation}
\label{sec:approx}

Using the area of the convex hull
as an upper bound can also be used for approximating
\textsc{Max-Area}.

\begin{theorem}[\cite{f-gtsp-92}]
\label{approx}
Let $P$ be a set of $n$ points in the plane. We can determine
a simple polygon $\mathcal{P}$ on $P$ that has area larger than
$\frac{1}{2}\AR(P)$,
where $\AR(P)$ denotes the area of $\hull(P)$. This can
be done in time $O(n\log n)$.
\end{theorem}

\begin{figure}[h]
        \centering
        \includegraphics[height=0.3\textwidth]{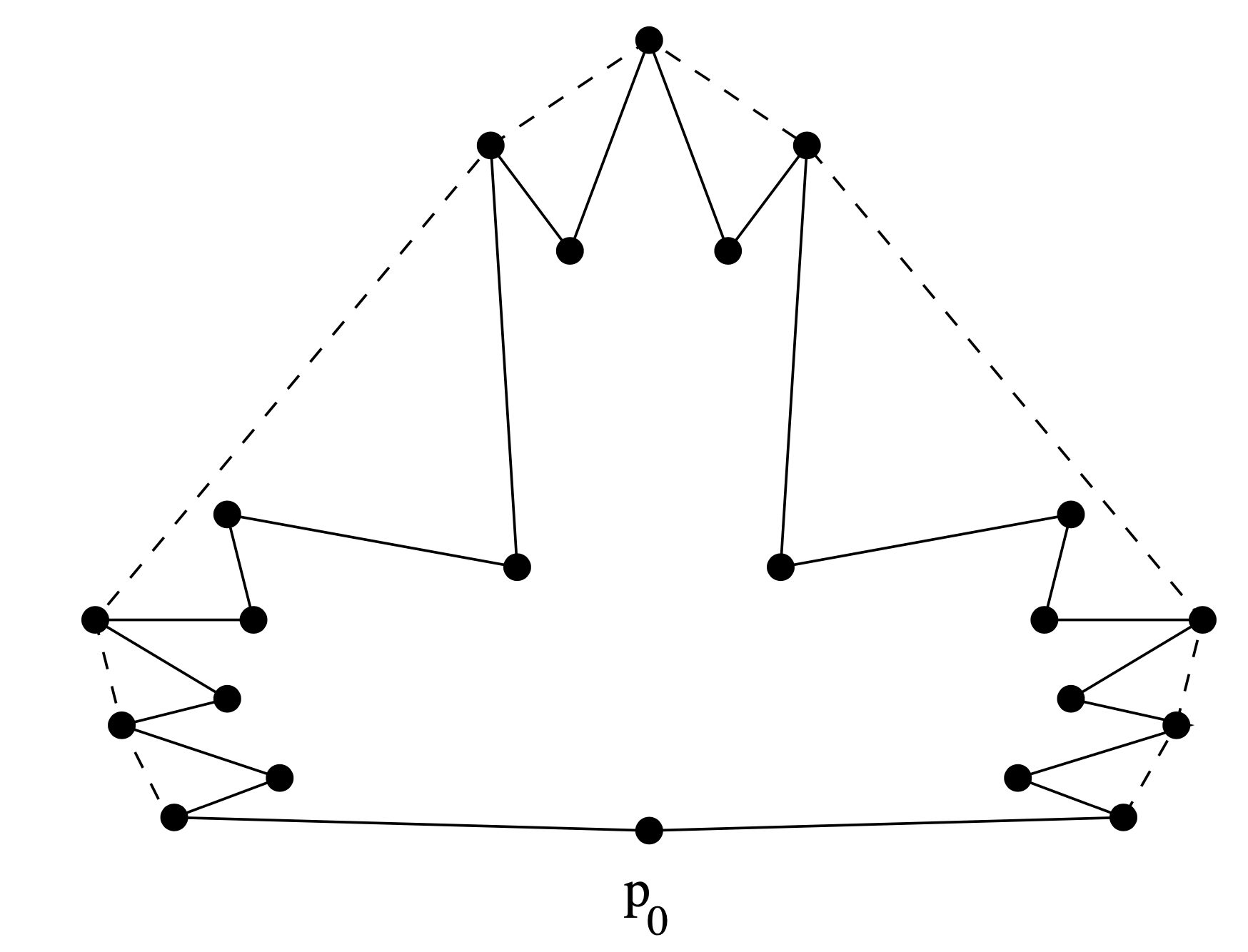}
        \includegraphics[height=0.3\textwidth]{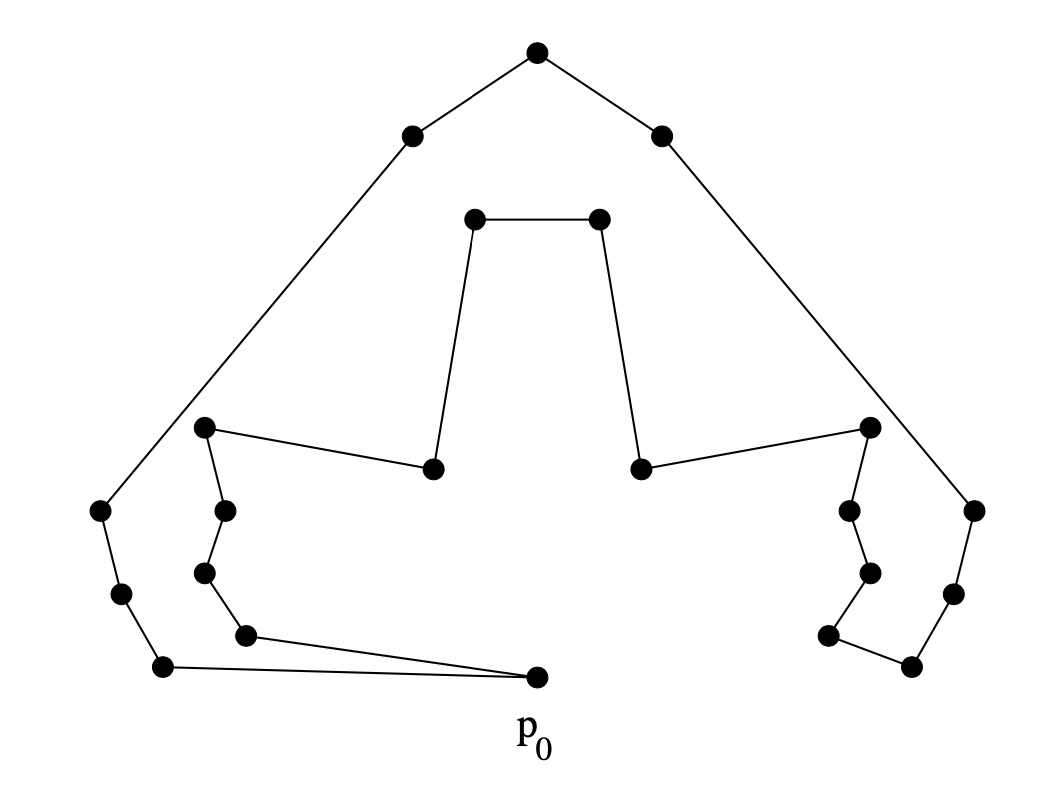}
        \caption{A $\frac{1}{2}$-approximation for \textsc{Max-Area}: A star-shaped
polygon $\mathcal{P}_1$ around a hull point $p_0$ (left) and a second simple
polygon $\mathcal{P}_2$ covering the rest of the convex hull.}
        \label{fig:approx}
\end{figure}

\begin{proof}
Let $p_0$ be a point on the convex hull of $P$.
In time $O(n\log n)$,
sort the points $p_i$ of $P$ by the slope
of the lines $l(p_0,p_i)$, such that the neighbors of $p_0$
on the convex hull are the first and the last point, respectively.
If there is a set of points for which the slope is the same, break the
tie by ordering them in increasing distance from $p_0$, except when
those points have the smallest of all slopes, in which case we take
them in order of decreasing distance from $p_0$.
Connecting the points $p_i$ in this order yields a simple
polygon $\mathcal{P}_1$ on $P$.

If $\AR(\mathcal{P}_1) > \frac{1}{2} \AR(P)$, we are done.
Suppose this is not the case. Then the set $\mathcal{Q}
:= \hull(P)\setminus \mathcal{P}_1$ has area  at least $\frac{1}{2}\AR(P)$.
Now it is not hard to see that there is a simple polygon $\mathcal{P}_2$
that contains $\mathcal{Q}$, implying the claim.
\end{proof}

\begin{figure}[h]
        \centering
        \includegraphics[height=0.3\textwidth]{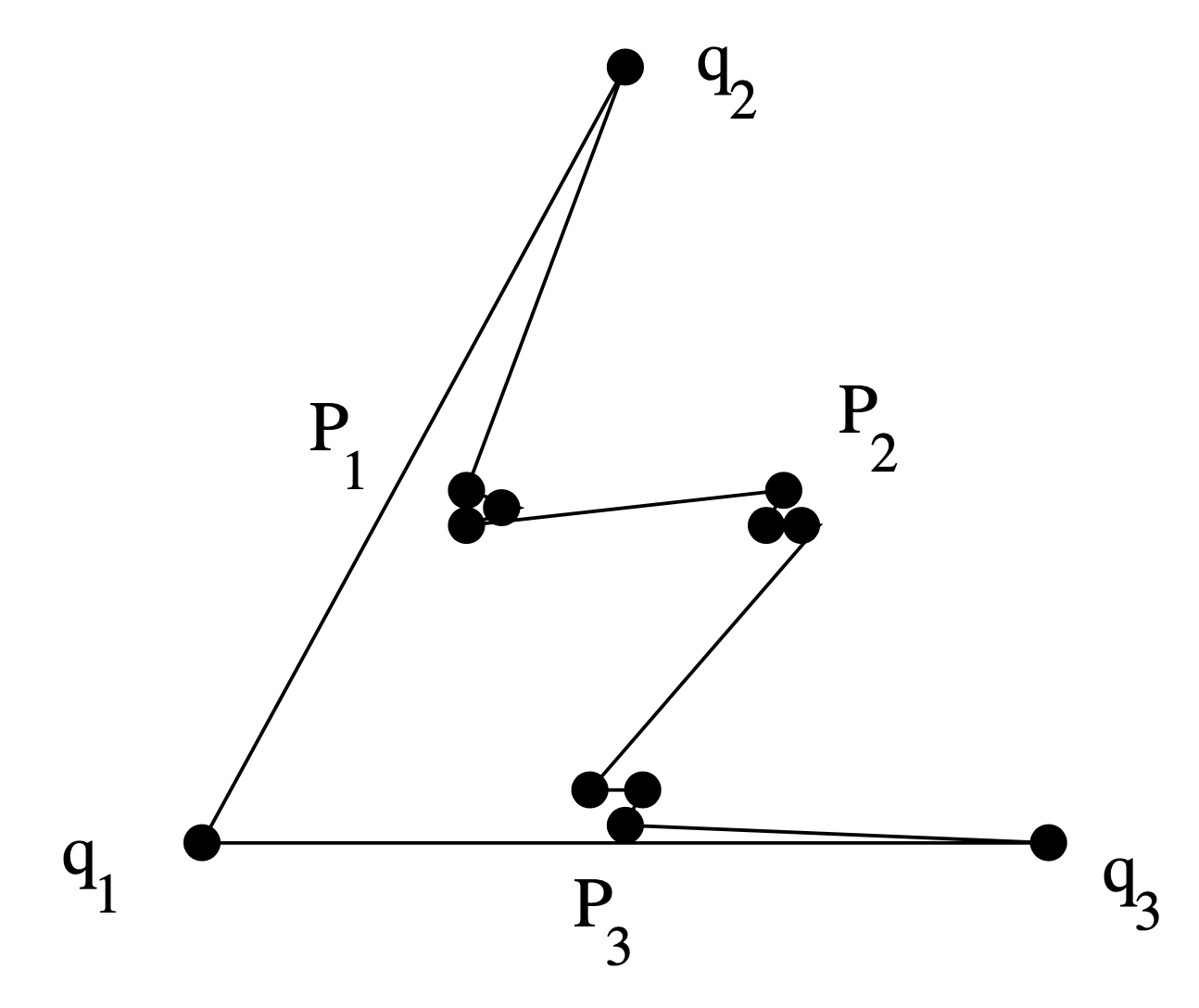}
        \includegraphics[height=0.3\textwidth]{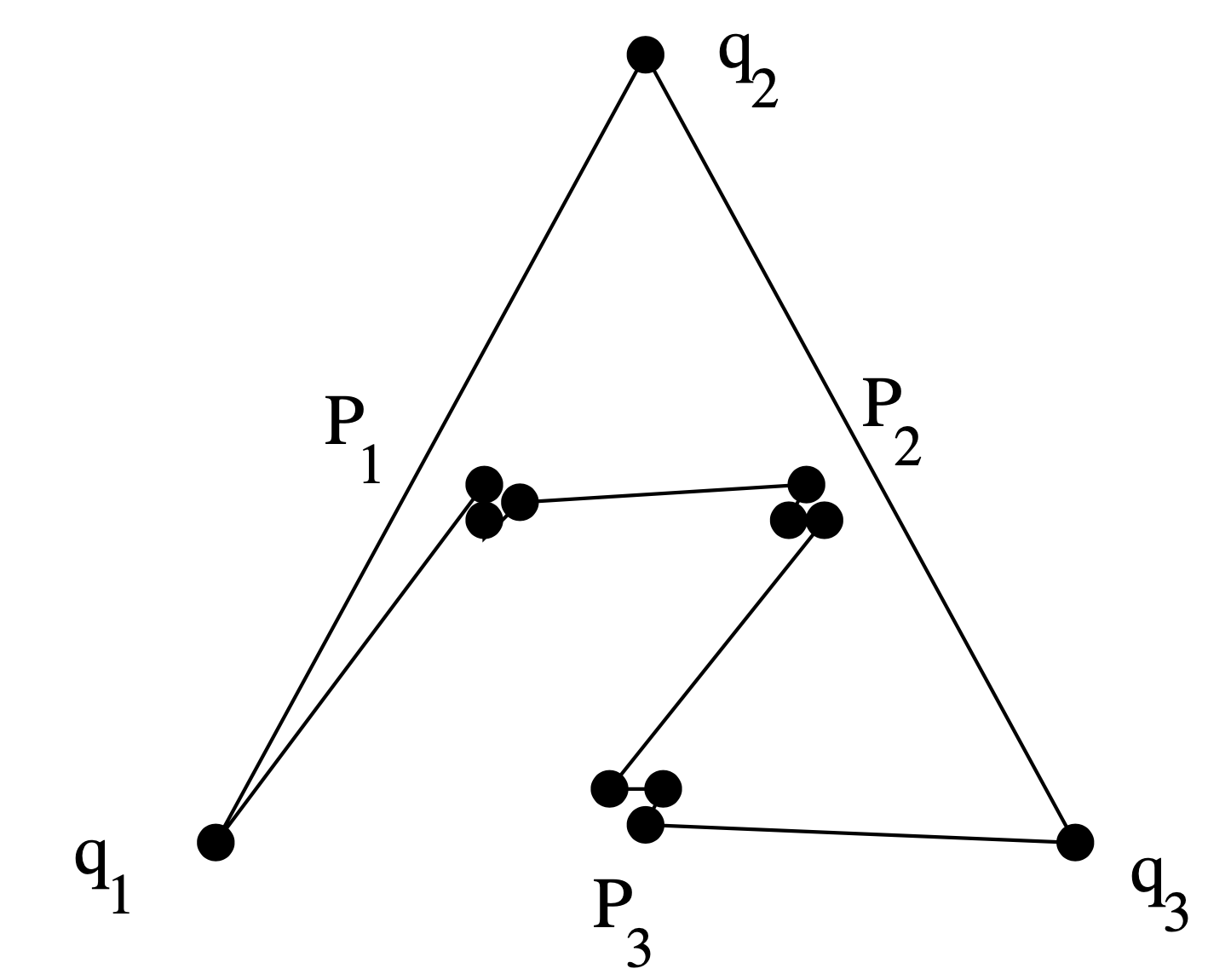}
        \caption{The approximation factor of $\frac{1}{2}$ for the
algorithm is tight.}
        \label{fig:tight}
\end{figure}

As shown in Figure~\ref{fig:tight},
$\frac{1}{2}$ is a tight bound for this
approach, even if all possible choices for $p_0$ are tested.
Moreover, it was shown in~\cite{f-gtsp-92}
that it is $\mathcal{NP}$-complete to decide whether there is
a simple polygon that contains strictly 
more than $\frac{2}{3}$ of the area of the convex hull;

At this time, no constant-factor approximation algorithm for \textsc{Min-Area}
is known, hinting at a higher level of difficulty of the minimization problem. 
It may well be the case that no such approximation can be computed
in polynomial time.

\section{Contest and Outcomes}
\label{sec:contest}

The 2019 Challenge was well received, with 28
teams from all over the world and 
a range of different scientific areas competing; participation was open
to anyone.  The contest itself was run through a dedicated server at TU Braunschweig,
hosted at \url{https://cgshop.ibr.cs.tu-bs.de/competition/cg-shop-2019/}.
It opened on February 28, 2019, and closed on May 31, 2019.

\subsection{Instances}
The contest started with a total of 247 benchmark instances, as follows.
Each of these instances consisted of $n$ points in the plane with integer coordinates.
For $n\in\{10$, $15$, $20$, $25$, $30$, $35$, $40$, $45$, $50$, $60$, $70$, 
$80$, $90$, $100$, $200$, $300$, $400$, $500$, $600$, $700$, $800$, $900$, $1000$, $2000$, $3000$, $4000$, $5000$, $6000$, $7000$, $8000$,
$9000, 10000, 20000, 30000, 40000, 50000, 60000, 70000, 80000$, $90000, 100000\}$,
there were six instances each.  In addition, there is one instance of size 
$n=1000000$.

The instances were of four different types; see Fekete et al.~\cite{dhh+-cnpmp-17} for a more detailed description 
of how to generate benchmark instances through illumination maps.

\begin{itemize}
\item
\textbf{uniform:} uniformly distributed at random from a square
\item
\textbf{edge:} randomly generated according to the distribution of the rate of change (the ``edges'') from various images
\item
\textbf{illumination:} randomly generated according to the distribution of brightness of an image (such as an illumination map)
\item
\textbf{orthogonally collinear points:} randomly generated on an integral grid, resulting in a large number of collinear points (similar to printed circuit boards and distorted blueprints).
.
\end{itemize}

\subsection{Evaluation}
The comparison between different teams was based on an overall \emph{score}.
For each instance, this score is the ratio given by the achieved area divided
by the area of the convex hull; thus, the score is a number between 0 and 1. 
The total score achieved by each team was the sum of all 247 individual instance scores. 
Feasibility of submitted solutions was checked at the time of upload;
for instances without a feasible solution, a default score of 1 (for minimization) or 0
(for maximization) was used. For multiple submissions by the same team,
only the best feasible solution submitted was used to compute the score. 
In case of ties, the tiebreaker was set to be the date/time a specific score was obtained. 
This turned out not to be necessary.

\subsection{Results}
\begin{figure}[]
        \begin{center}
          \includegraphics[width=.95\textwidth]{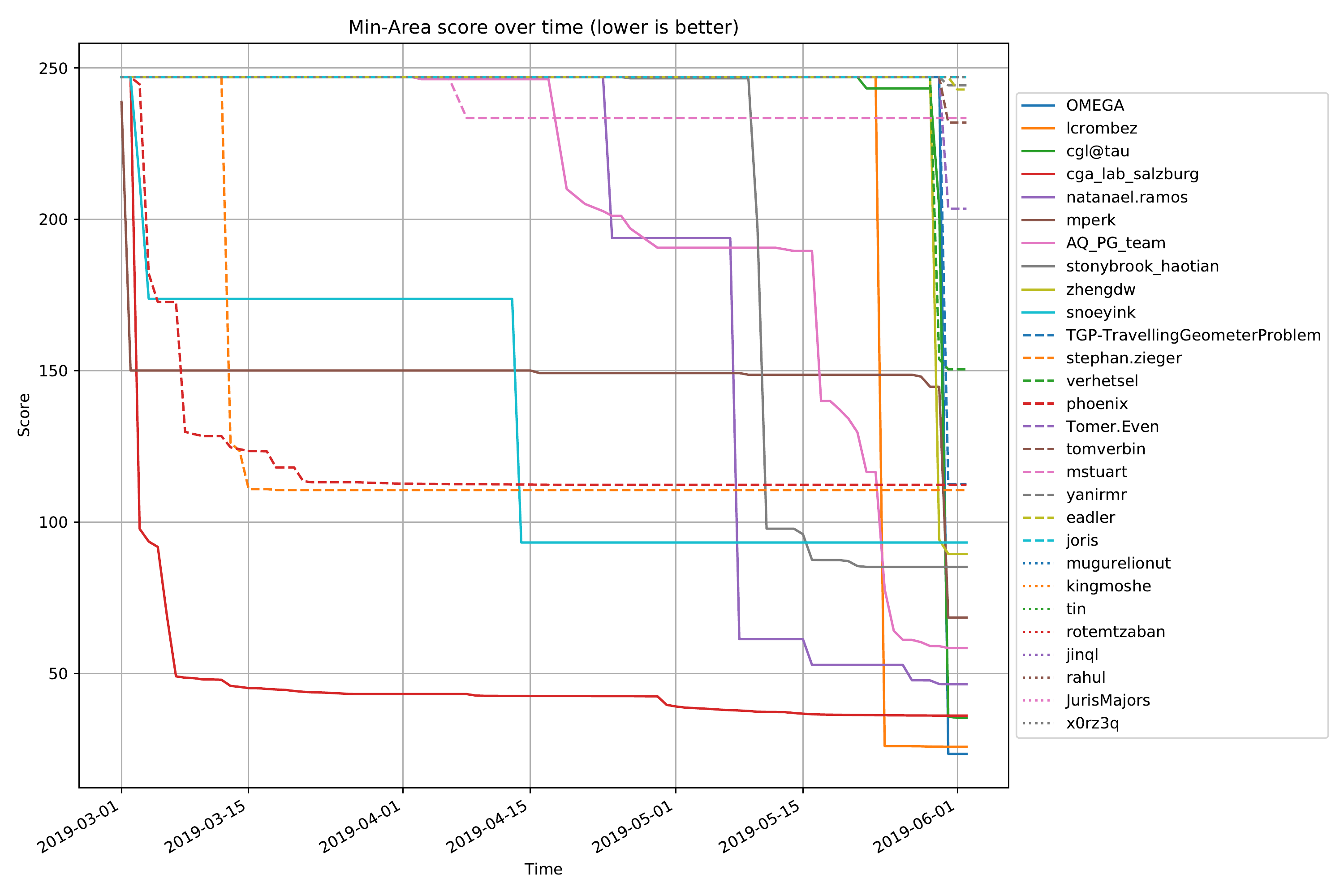}
        \end{center}
                \caption{Total score over time for \textsc{Min-Area}.}
        \label{fig:min}
\end{figure}

\begin{figure}[]
        \begin{center}
          \includegraphics[width=.95\textwidth]{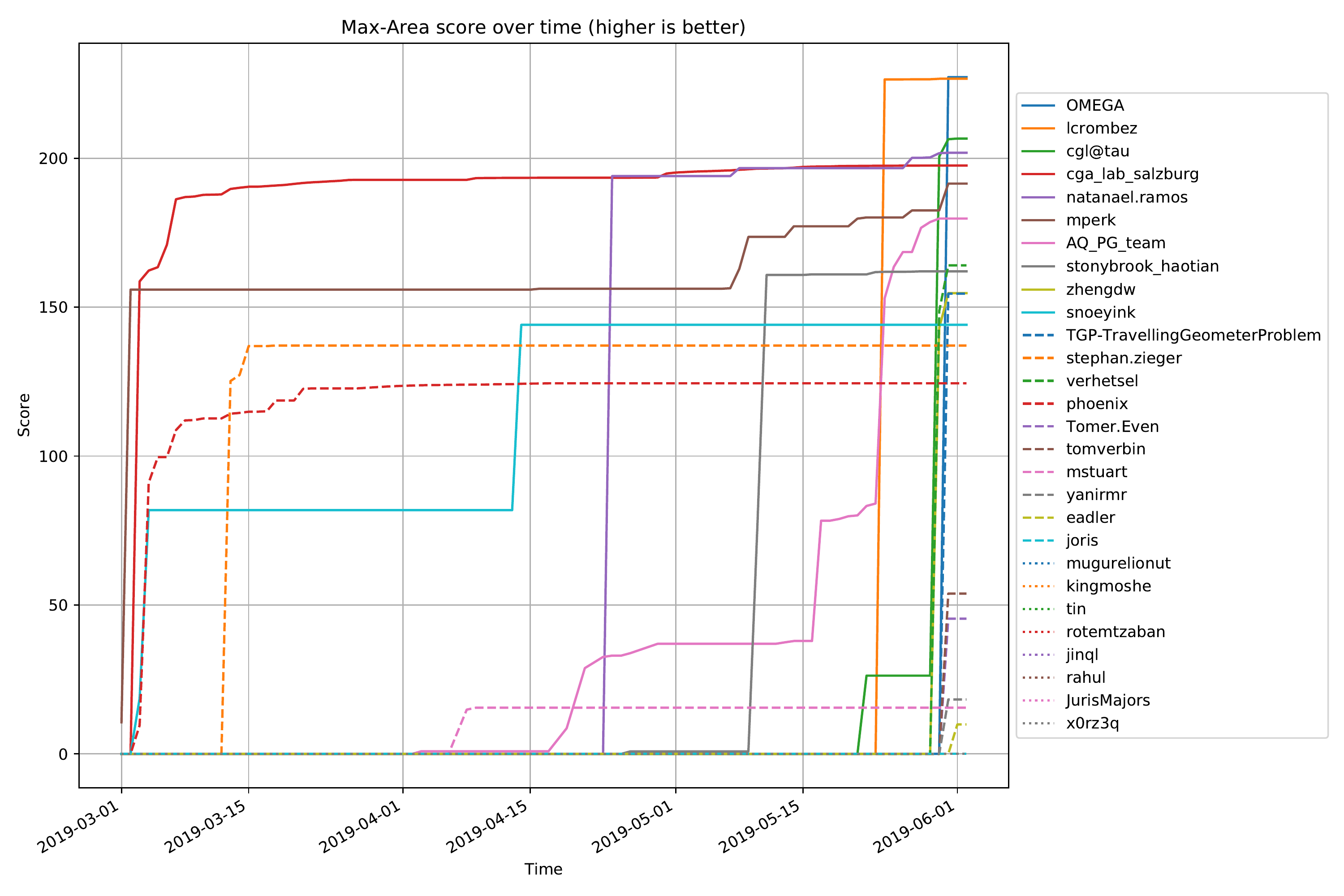}
        \end{center}
                \caption{Total score over time for \textsc{Max-Area}.}
        \label{fig:max}
\end{figure}

In the end, the top 10 in the leaderboard
looked as shown in Table~\ref{tab:top10}; note that according to the scoring function,
a lower score is better for \textsc{Min-Area}, while a higher score is better for \textsc{Max-Area}.
The progress over time of each team's score can be seen in Figure~\ref{fig:min} (for \textsc{Min-Area})
and Figure~\ref{fig:max} (for \textsc{Max-Area}).

\begin{table}[h!]
  \begin{center}
    \caption{The top of the final leaderboard. Shown are the scores in both
categories, along with the achieved average percentages of the convex hull of
points. Teams CGA and UNICAMP were overall tied, according to different
positions for \textsc{Min-Area} and \textsc{Max-Area}, as were mperk and
AQ\_PG. The teams mperk (Michael Perk, TU Braunschweig) and zhengdw (David
Zheng, University of British Columbia) were both individual, first-year Masters students.}
    \label{tab:top10}
    \begin{tabular}{|r|l|r|r|rr|rr|} 
      \hline
      \textbf{Rk.} & \textbf{Team} & \textbf{Score (Min)} & \textbf{Score (Max)} 	& \textbf{\# best} &\textbf{(unique)}  & \textbf{\# best} &\textbf{(unique)}\\
      			& 		&		       & 			& \textbf{Min} &\textbf{sols.} & \textbf{Max} &\textbf{sols.}\\
      \hline
      1 & OMEGA/Mulhouse (FR)		& 23.393 (\ \ 9.47\%) 	& 227.247 (92.00\%) 		& 223 &(181) & 184 &(138) \\
      2 & lcrombez/Clermont (FR)	& 25.751 (10.43\%)	& 226.691 (91.78\%)		& 66  &(23)  & 109 &(4) \\
      3 & cgl@tau/Tel Aviv (IS) 	& 35.289 (14.29\%)	& 206.612 (83.65\%)		& 13  &(0)   & 12  &(0) \\
      4 & CGA/Salzburg (AU)		& 36.069 (14.60\%)	& 197.568 (79.99\%)		& 26  &(0)   & 23  &(0) \\
      4 & UNICAMP/Campinas (BR)		& 46.432 (18.80\%)	& 201.839 (81.72\%)		&  3  &(0)   & 3   &(0) \\
      6 & mperk/Braunschweig (GE)	& 68.431 (27.70\%)	& 191.483 (77.52\%)		& 24  &(0)   & 23  &(0) \\
      6 & L'Aquila-Perugia (IT)		& 57.373 (23.23\%)      & 179.752 (72.77\%)             & 19  &(0)   & 18  &(0) \\
      8 & Stony Brook (US)   		& 85.179 (34.49\%)      & 162.031 (65.60\%)             &  1  &(0)   & 1   &(0) \\
      9 & zhengdw (CA)   		& 89.437 (36.21\%)      & 154.723 (62.64\%)             &  0  &(0)   & 1   &(0) \\
      10 & TGP/Eindhoven (NL)		& 112.561 (45.57\%)     & 154.548 (62.57\%)             &  18 &(0)   & 26  &(0) \\
      \hline
    \end{tabular}
  \end{center}
\end{table}



The top 5 finishers were invited for contributions to this special issue, as follows.

\begin{enumerate}
\item Team OMEGA/Mulhouse (France): Julien Lepagnot, Laurent Moalic, Dominique Schmitt~\cite{area-omega}
\item Team lcrombez/Clermont Auvergne(France): Lo\"ic Crombez, Guilherme D. da Fonseca, Yan Gerard~\cite{area-crombez}
\item Team cgl@tau/Tel Aviv (Israel): Nir Goren, Efi Fogel, Dan Halperin~\cite{area-tau}
\item Team CGA/Salzburg (France): G\"unther Eder, Martin Held, Steinthor Jasonarson, Philipp Mayer, Peter Palfrader~\cite{area-salzburg}.
\item Team UNICAMP/Campinas (Brazil): Natanael Ramos, Rai Caetan de Jesus, Pedro de Rezende, Cid de Souza, Fabio Luiz Usberti~\cite{area-campinas}
\end{enumerate}

Figure~\ref{fig:min} shows the development of total scores over time for \textsc{Min-Area}
and all teams; it can be seen that OMEGA passed lcrombez shortly before the deadline,
with cgl@tau just squeezing by CGA. A similar and even tighter
outcome between OMEGA and lcrombez for \textsc{Max-Area} can be seen in Figure~\ref{fig:max};
for that problem, cgl@tau also placed third, but UNICAMP managed to beat out CGA for fourth place.
Thus, the ranking was consistent for both \textsc{Min-Area} and \textsc{Max-Area}, except for Campinas doing better than
Salzburg in \textsc{Max-Area}, so they both shared fourth place.
All five teams engineered their solutions with the use of a variety of specific tools.
Details of their methods and the engineering decisions they made are given in their respective papers.

\begin{figure}[]
        \begin{center}
          \includegraphics[width=.95\textwidth]{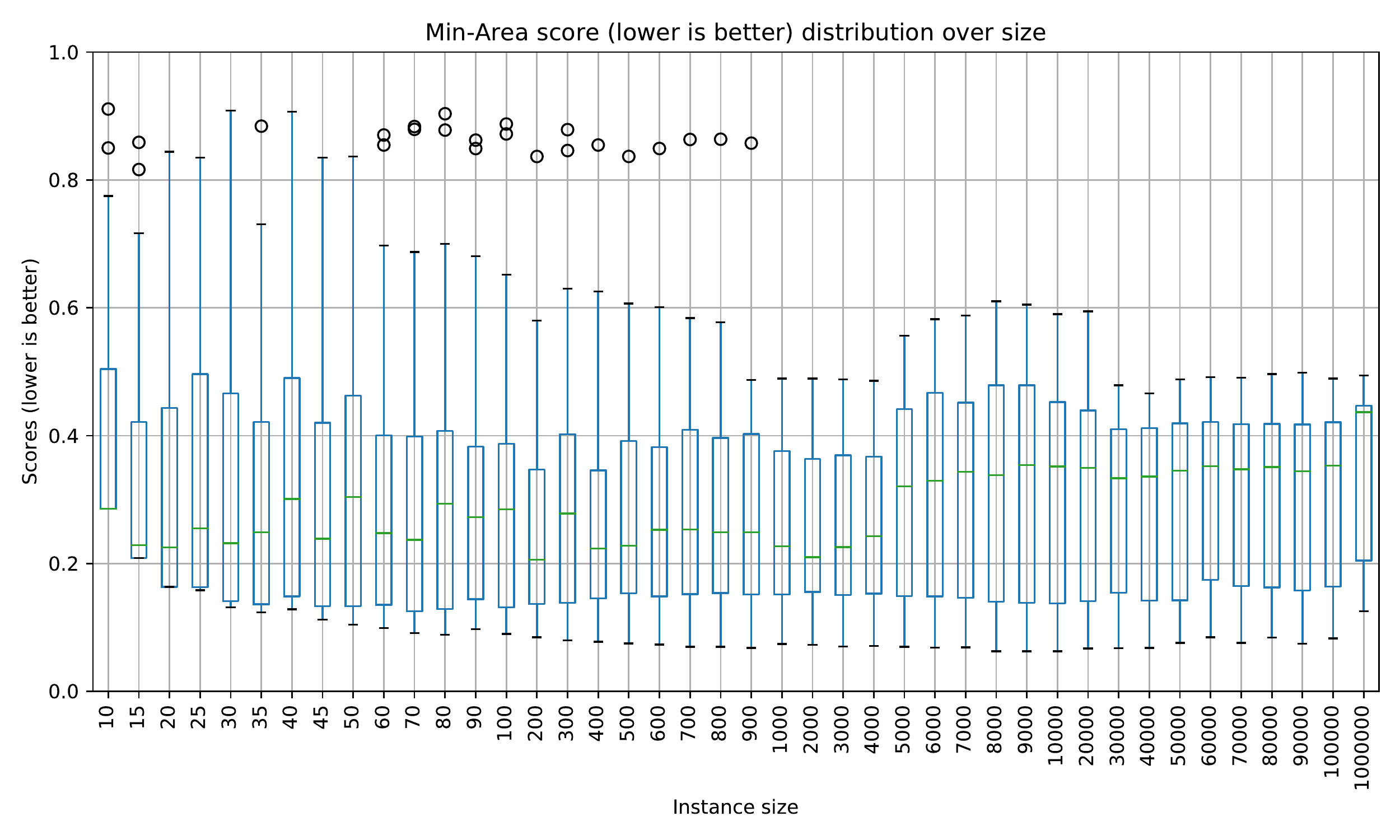}
        \end{center}
                \caption{Distribution of best scores over all \textsc{Min-Area} instances. The green line shows the median. The boxes show the quartiles and the whiskers the min and max (except possibly outliers as additional circles). If there were multiple instances per size, the mean score for those is used.}
        \label{fig:minscores}
\end{figure}

\begin{figure}[]
        \begin{center}
          \includegraphics[width=.95\textwidth]{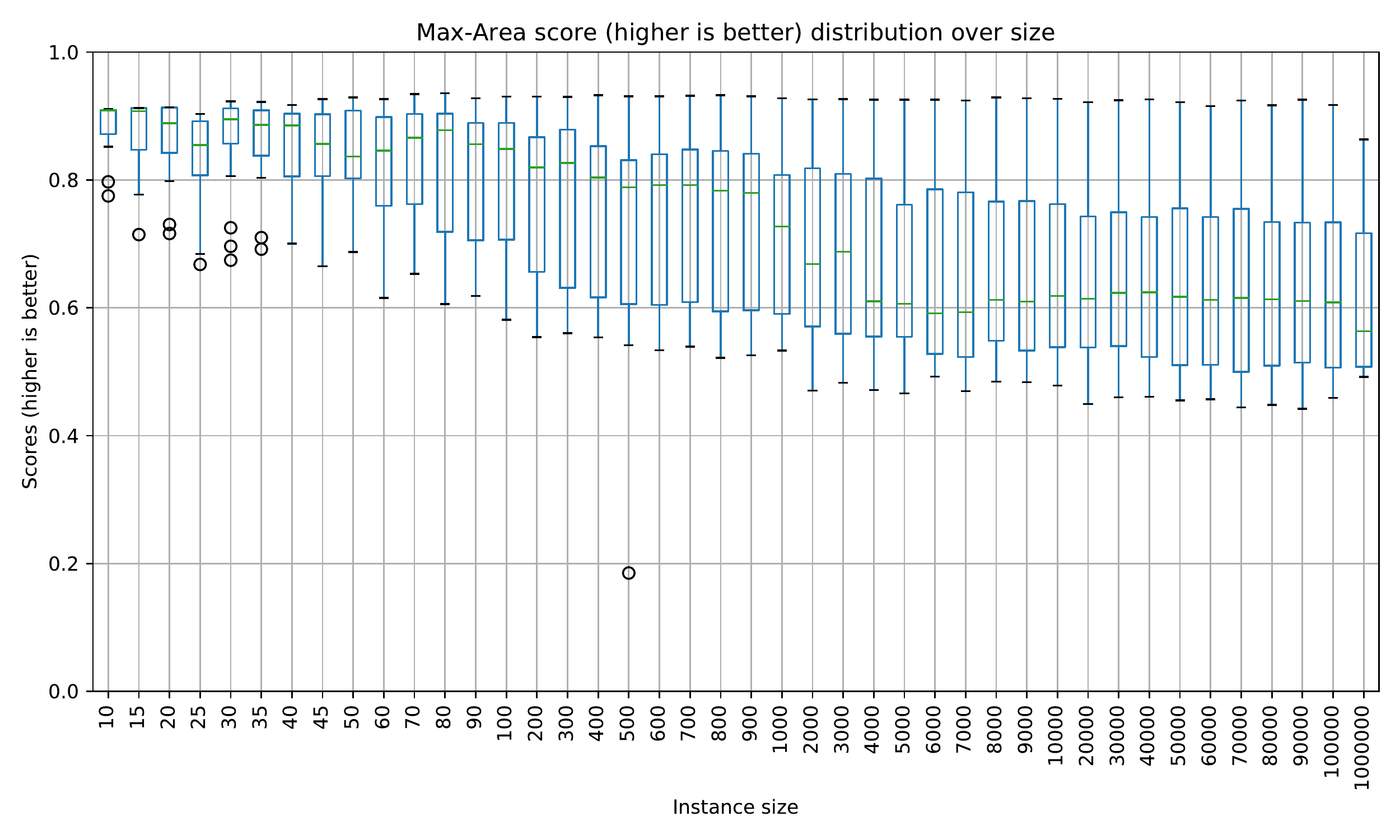}
        \end{center}
                \caption{Distribution of best scores over all \textsc{Max-Area} instances. The green line shows the median. The boxes show the quartiles and the whiskers the min and max (except possibly outliers as additional circles). If there where multiple instances per size, the mean score for those is used.}
        \label{fig:maxscores}
\end{figure}

Figure~\ref{fig:minscores} shows the spread of results for all \textsc{Min-Area} instances; it can be seen
that there is a strong deviation over all instance sizes, even for the small ones.
This is surprising as one would expect simple heuristics to perform reasonably well for small instances;
however, it appears that even for very small instances with $10$ points of
\textsc{Min-Area}, more advanced ideas are necessary to achieve good results.
(See the contribution by Fekete et al.~\cite{area-exact} for a detailed study of exact methods.)
At around $5000$ points, the mean score drops visibly; the best teams are able to obtain nearly the same score for all instances sizes above $50$ points.
A similar overview for \textsc{Max-Area} is given in Figure~\ref{fig:maxscores}.
Here the deviation is very small for the small instances, showing that all teams where able to obtain reasonably good solutions for small instances.
The deviation increases with the problem size until around $300$ points, after which it remains homogenous but smaller than for \textsc{Min-Area}.
Like for \textsc{Min-Area}, the best teams obtained nearly equal scores for all instances sizes,
 while the mean score drops visibly after $900$ points.

\section{Conclusions}
The 2019 CG Challenge motivated a considerable number of teams to engage in intensive optimization studies.
This success has not only led to practical progress on the problem of area optimization,
but also turned the CG Challenge into a continuing feature of CGWeek, spawning considerable work
through the 2020 Challenge problem (Minimum Convex Partition) and the 2021 problem (Coordinated
Motion Planning). This promises to motivate further work
on the involved problems, as well as other practical geometric optimization work.
We are confident that this will further strengthen the bridges between optimization theory and
practical algorithm engineering.

\section*{Acknowledgment}
This work was supported by DFG project ``Computational Geometry: Solving Hard Optimization Problems'' (CG:SHOP), FE 407/21-1.

\bibliographystyle{ACM-Reference-Format}
\bibliography{lit}

\end{document}